\documentclass[aps,pre,twocolumn,showpacs,superscriptaddress,longbilography]{revtex4-1}

\usepackage{amsbsy,amssymb,amsmath,bm}
\usepackage{graphicx}
\usepackage{braket}
\usepackage{float}
\usepackage{textcomp}
\usepackage{color}
\usepackage[colorlinks=true,linkcolor=red]{hyperref}
\usepackage[sort&compress]{natbib}

\usepackage{soul}

\begin{document}

\title {Almost compact moving breathers with fine-tuned discrete time quantum walks }

\author{I. Vakulchyk}
\affiliation{Center for Theoretical Physics of Complex Systems, Institute for Basic Science (IBS), Daejeon 34051, Republic of Korea}
\affiliation{Basic Science Program, Korea University of Science and Technology (UST), Daejeon 34113, Republic of Korea}

\author{M. V. Fistul}
\affiliation{Center for Theoretical Physics of Complex Systems, Institute for Basic Science (IBS), Daejeon 34051, Republic of Korea}
\affiliation{Russian Quantum Center, National University of Science and Technology "MISIS", Moscow 119049 Russia}

\author{Y. Zolotaryuk}
\affiliation{Center for Theoretical Physics of Complex Systems, Institute for Basic Science (IBS), Daejeon 34051, Republic of Korea}
\affiliation{Bogolyubov Institute for Theoretical Physics
National Academy of Sciences of Ukraine, Kiev 03143 
Ukraine}

\author{S. Flach}
\affiliation{Center for Theoretical Physics of Complex Systems, Institute for Basic Science (IBS), Daejeon 34051, Republic of Korea}

\date{\today}
\begin{abstract}
Discrete time quantum walks are unitary maps defined on the Hilbert space of coupled two-level systems.
We study the dynamics of excitations in a nonlinear discrete time quantum walk, whose fine-tuned linear counterpart has a flat band structure. 
The linear counterpart is, therefore, lacking transport, with exact solutions being compactly localized. A solitary entity of the nonlinear walk moving at velocity $v$ would 
therefore not
suffer from resonances with small amplitude plane waves with identical phase velocity, due to the absence of the latter. That solitary excitation would also
have to be localized stronger than exponential, due to the absence of a linear dispersion.
We report on the existence of a set of stationary and moving breathers with almost compact superexponential spatial tails. 
At the 
limit of the largest velocity $v=1$ the moving breather turns into a completely compact bullet. 
\end{abstract}

\maketitle
\textbf{In this work, we utilize a nonlinear generalization of a discrete time quantum walk (DTQW) which is widely used in the field of quantum computing including experimental realizations. This unitary map toolbox allows to study numerically
the dynamics of solitary type excitations in discrete lattices with great efficiency. The linear DTQW is fine-tuned to a dispersion relation with two flat bands which
inhibit transport and allow for compact localized states. Nonlinearity leads 
to the appearance of stable moving and stationary nonlinear excitations which are super-exponentially localized. 
The system further allows for fully compact bullet excitations moving with the maximum velocity, which may also form an intriguing interacting gas with
unusual scattering properties.
}

\section{Introduction}
Discrete breathers (DBs) \cite{flach1998discrete,Flach:2008} are generic time-periodic and spatially localized solutions to broad classes of nonlinear Hamiltonian network equations \cite{ma94n}. Discrete breathers have been observed and studied in a fascinating and broad setting of physical realizations which cover several decades
of temporal and spatial scales (see [\onlinecite{Flach:2008}] for a detailed review).
If the band structure of the linear part of the Hamiltonian network is not degenerated, then DBs are localized in space either following an exponential decay (for analytical
band structure functions) or an algebraic one (for non-analytical ones) \cite{f98pre}. If the band structure is degenerated and consists of flat band(s) only, DBs localize
superexponentially fast \cite{deft01PRE}. For specific local symmetries of flat bands, DBs can even maintain the compactness of the corresponding compact localized eigenstates of the linear Hamiltonian network equations \cite{doi:10.1063/1.5041434}.

The generic appearance of DBs in lattice structures comes at a price. The absence of a continuous translational invariance (which is replaced at best by some
discrete one) makes travel hard. Indeed, multiple attempts to obtain lossless traveling discrete breathers suffered from facing resonances between the
velocity $v$ of a moving DB candidate, and phase velocities of small amplitude plane waves \cite{Flach:2008}. These - usually unavoidable - resonances produce
nondecaying tails. Relief could be only obtained by choosing generalized discrete nonlinear Schr\"odinger equations. 
Their global phase (or simply gauge) symmetry 
allows finding stationary DBs supported by just one harmonics in time. Then tailless moving DBs could be obtained for a discrete and non-empty 
set of velocity values $v$ \cite{mckc06prl,PhysRevE.76.036603}. 
In contrast, systems with continuous translational invariance not
only allow for Galilean or Lorentz boosting of solitary excitations if their stationary parents exist but also permit 
the occurrence of completely compact moving
solitary excitations in nonlinear partial differential equations which 
have a missing linear dispersive part \cite{1751-8121-51-34-343001}.
Interestingly, G. James reported recently on an attempt to find traveling DBs in a strongly nonlinear discrete nonlinear Schr\"odinger chain with a missing linear dispersive
part \cite{James20170138}. However even in this case nondecaying tails were observed. 

In this work, we consider nonlinear generalizations of discrete time quantum walks (DTQW) as a promising way to overcome the above difficulties of Hamiltonian networks.
DTQW were introduced as quantum generalizations of classical random walks by Aharonov et al. \cite{PhysRevA.48.1687}.
The  DTQW  evolution  is given by a  (discrete)  sequence  of  unitary operators acting on a quantum state of a chain of
two-level systems in a high-dimensional Hilbert space \cite{kempe2003quantum,chandrashekar2007symmetries,vakulchyk2017anderson}.  DTQW exhibit quantum interference and superposition \cite{PhysRevA.48.1687},  
entanglement \cite{abal2006quantum},  two-body coupling  of  wave  functions \cite{omar2006quantum},  and Anderson  localization
\cite{crespi2013anderson,vakulchyk2017anderson}. DTQW experimental realizations were reported
with ion trap systems \cite{schmitz2009quantum}, quantum optical waveguides \cite{peruzzo2010quantum}, 
and nuclear magnetic resonance quantum computing \cite{du2003experimental}.
DTQW maps can be considered as generated by Floquet driven quantum Hamiltonians, without precise knowledge of the Hamiltonian. Through a fine-tuning of
the DTQW control parameters, it is straightforward to inhibit transport by making the complete band structure flat \cite{vakulchyk2017anderson}. 
Moreover, it is straightforward to include nonlinear terms while keeping the unitarity of a generalized nonlinear DTQW \cite{Vakulchyk:2018aa}.

By making use of direct numerical simulations and complimentary generalized Newton schemes we arrive at a plethora of nonlinear DTQW breathers.
The breathers have superexponential tails, can be stationary, but also moving with a dense set of velocities. Analytical results fit well with numerical observations.
The paper is organized as follows: In Section II we introduce the nonlinear DTQW. 
In Section III we study the resulting dynamics. Section IV provides with conclusions.

\section{The nonlinear discrete time quantum walk}
\label{II}
The dynamics of the nonlinear DTQW is evolving a two-component wave function $\hat \psi_n(t)=\{ \psi_{+,n}(t), \psi_{-,n}(t) \}$ in space and time. The nonlinear DTQW dynamics is controlled by coin operators $\hat U_n$ which are unitary
matrices in general determined by four site dependent angles \cite{vakulchyk2017anderson}. For our purposes we choose the simplest and generic version controlled by only one angle type:
\begin{equation}\label{coin_operator}
    \hat U_n 
     = \begin{pmatrix}
        \cos{\theta_n} &  \sin{\theta_n} \\
         -\sin{\theta_n} &  \cos{\theta_n}
        \end{pmatrix},
\end{equation}
where the angles $\theta_n=\theta+\lambda S_n$  are nonlinear functions of $\hat \psi_n$ with the norm density per site $S_n=(|\psi_{+,n}|^2+|\psi_{-,n}|^2)$.  
Including the shift operator (see Ref. [\onlinecite{vakulchyk2017anderson}] for details) we arrive at the nonlinear unitary map of the whole chain of two-level systems
(see also the schematic DTQW evolution in Fig.\ref{fig1}): 
\begin{figure}
\includegraphics[width=0.99\columnwidth]{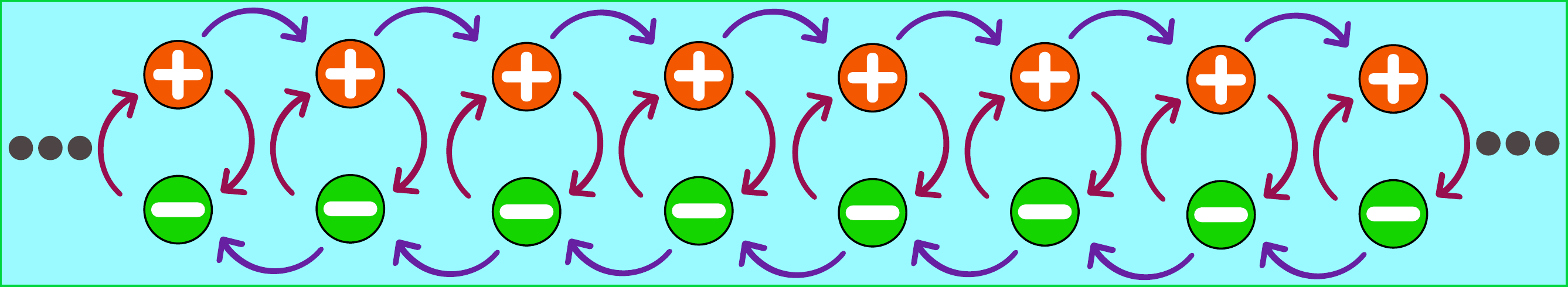}
\caption{(Color online) A schematic representation of a general discrete-time 
quantum walk. The vertical arrows indicate the quantum coin 
action within each two-level system, while the horizontal ones
show the action of the transfer operator. }
\label{fig1}
\end{figure}
\begin{widetext}
\begin{equation}
\psi_{\pm ,n}(t+1) = \cos \theta_{n\mp 1} \psi_{\pm ,(n \mp 1)} (t)
\pm \sin \theta_{n \mp 1} \psi_{\mp ,(n \mp 1)} (t) \;.
\label{DiffEquation}
\end{equation}
\end{widetext}

Note that for finite systems with $N$ sites periodic boundary conditions are to be applied.
The above evolution is unitary and preserves the total norm of the wave function $S=\sum_{n=1}^N S_n$.
Moreover, the model enjoys stroboscopic sublattice factorization (SSF), i.e. the evolution on even and odd sites decouples locally if iterated over two time units. Indeed, it follows from Eqs. (\ref{DiffEquation}) that $\hat \psi_{n}(t+2) =F\left( \hat \psi_{n-2}(t),\hat \psi_{n}(t),\hat \psi_{n+2}(t) \right)$. For even $N$ it follows that the dynamics on even and odd sites decouples completely, similar to the
two distinct sides of a M\"obius band with zero twists. Instead for odd $N$ the even and odd site dynamics is globally coupled: any excitation on an even site will explore all even sites until it reaches the boundary where it turns odd and vice versa. This is similar to a M\"obius band with a nontrivial twist.

In the linear regime $\lambda=0$ the coin operators turn identical and space-index independent. That allows to seek solutions in the form of plane waves  $\hat \psi_n(t)=\hat \psi \exp [i(\omega t + kn)] $. 
The corresponding dispersion relation reads \cite{vakulchyk2017anderson}
\begin{equation} \label{Dispeq}
    \cos(\omega) = \cos(\theta) \cos(k).
\end{equation}
We note that for the particular choice
$\theta=\pi/2$ the resulting band structure is composed of two flat bands $\omega=\pm \pi/2$. 
In this case, compact localized eigenstates exist  \cite{vakulchyk2017anderson}, and the quantum walk dynamics is quenched resulting in the halt of any
propagating wave. Adding back the nonlinear terms with $\lambda \neq 0$, any possible observed transport will be entirely due to the nonlinear terms which lead
to an interaction between the compact localized states.
For the remaining part of this work, we will use $\theta=\pi/2$ only.

\section{Discrete breathers}
\label{III}

\subsection{Numerical observations}

\begin{figure}
  \includegraphics[width=0.99\columnwidth]{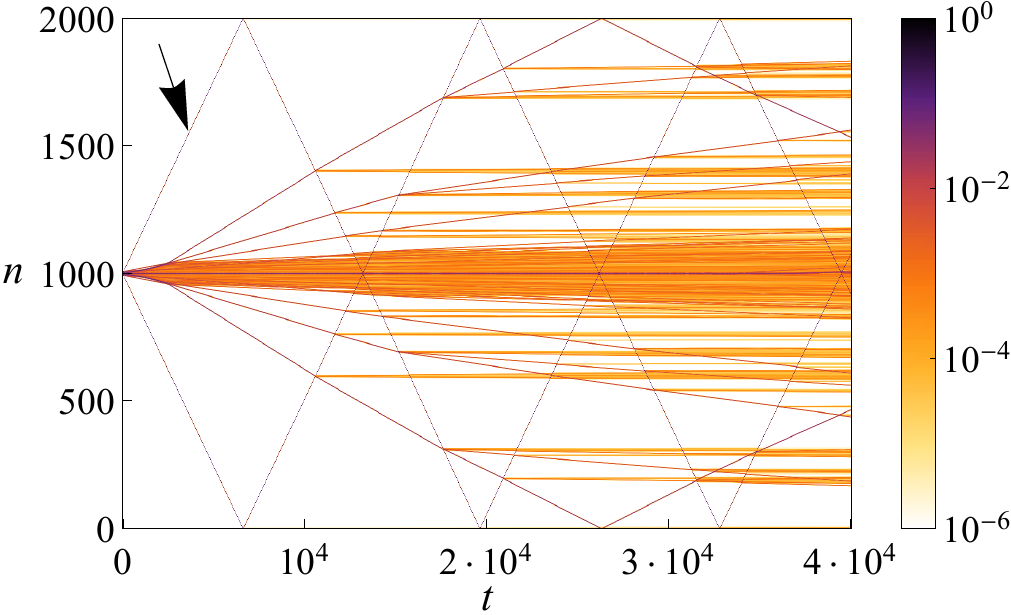}
    \caption{(Color online) Typical temporal and spatial pattern of  the nonlinear DTQW dynamics. Norm density $S_n$ is plotted versus time and space with color coding on a logarithmic scale.
$\lambda=0.1$,  with initial condition $\hat \psi_n(0)=\frac{\delta_{n,n_0}}{\sqrt{2}} \{1, 1 \}$ and $n_0=1000$. The arrow indicates the moving solitary excitation analyzed in greater detail in Fig.\ref{fig3}. }
    \label{fig2}
\end{figure}

We study the evolution of a single-site initial condition $\hat \psi_n(0)=\frac{\delta_{n,n_0}}{\sqrt{2}} \{1, 1 \}$ with the wave function norm $S=1$, launched
at site $n_0$. 
We use a system with size $N=2000$ and periodic boundary conditions, and evolve up to time $t=10^4$.
The resulting evolution of the norm density distribution is shown in Fig.\ref{fig2} for $\lambda=0.1$.
A part of the excitation remains in a relatively narrow core region which spreads, albeit very slowly. Our main observation is that the core
emits solitary type excitations at various times, which then continue to travel separately at various velocities. These objects are obvious candidates for traveling
discrete breathers. In the following, we will analyze them in detail.

\subsection{Cut and Paste}
Since the moving solitary type excitations are well separated in space, we apply a 'Cut and Paste' procedure. We (i) evolve the system
up to an appropriate cutting time $t_c$, (ii) identify the position $n_c$ of the core of a single isolated object, (iii) obtain the distance $l_c$ from the core at which the wave function amplitudes decay to noise levels $\sim 10^{-3}$, and (iv) zero the wave function amplitudes for all sites $n < n_c-l_c$ and $n>n_c+l_c$.
With a trivial re-shifting of time and space coordinates, we continue the evolution of the single moving object and its analysis.
In particular, we repeat the 'Cut and Paste' procedure several times, in order to allow the excitation to converge to a state which is almost radiationless, i.e. which is
not leaving weakly excited sites behind. In order to get a real-valued object, we take an absolute value of each component of $\hat{\psi}$ and repeat the procedure.

\begin{figure}
 \includegraphics[width=0.9\columnwidth]{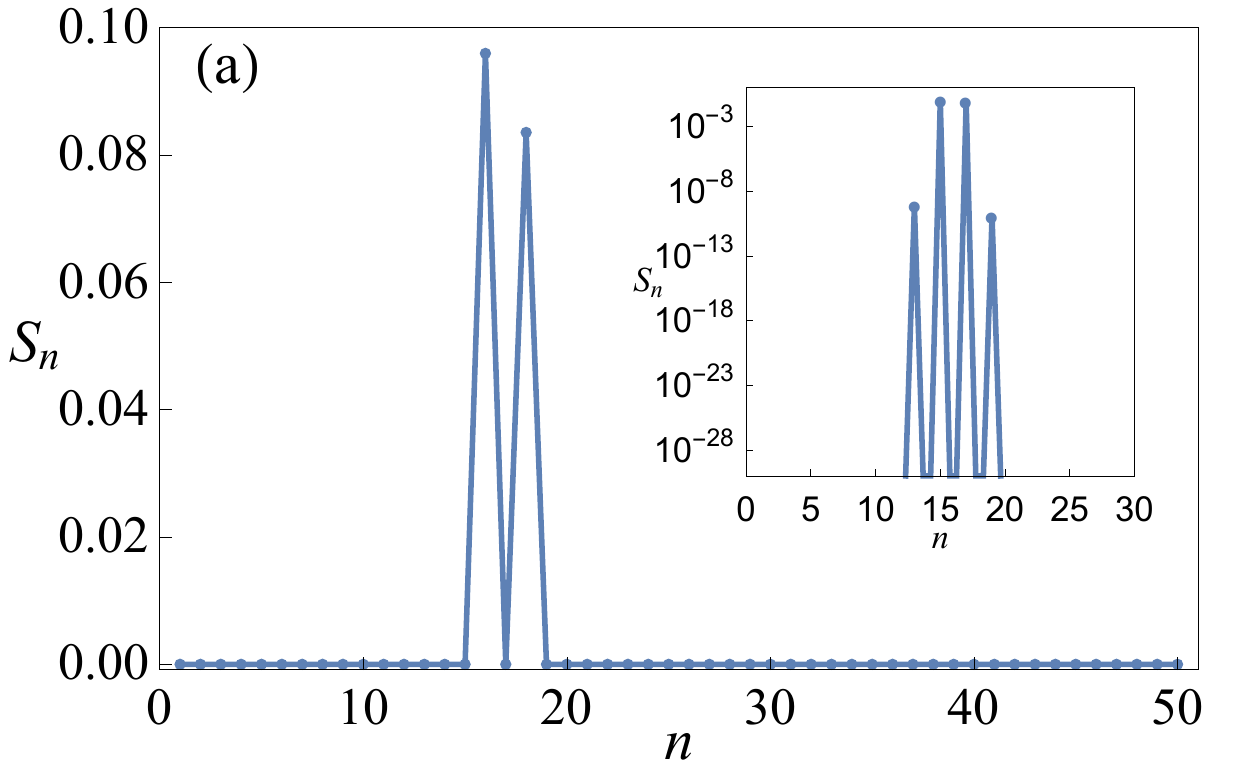}
  \includegraphics[width=0.9\columnwidth]{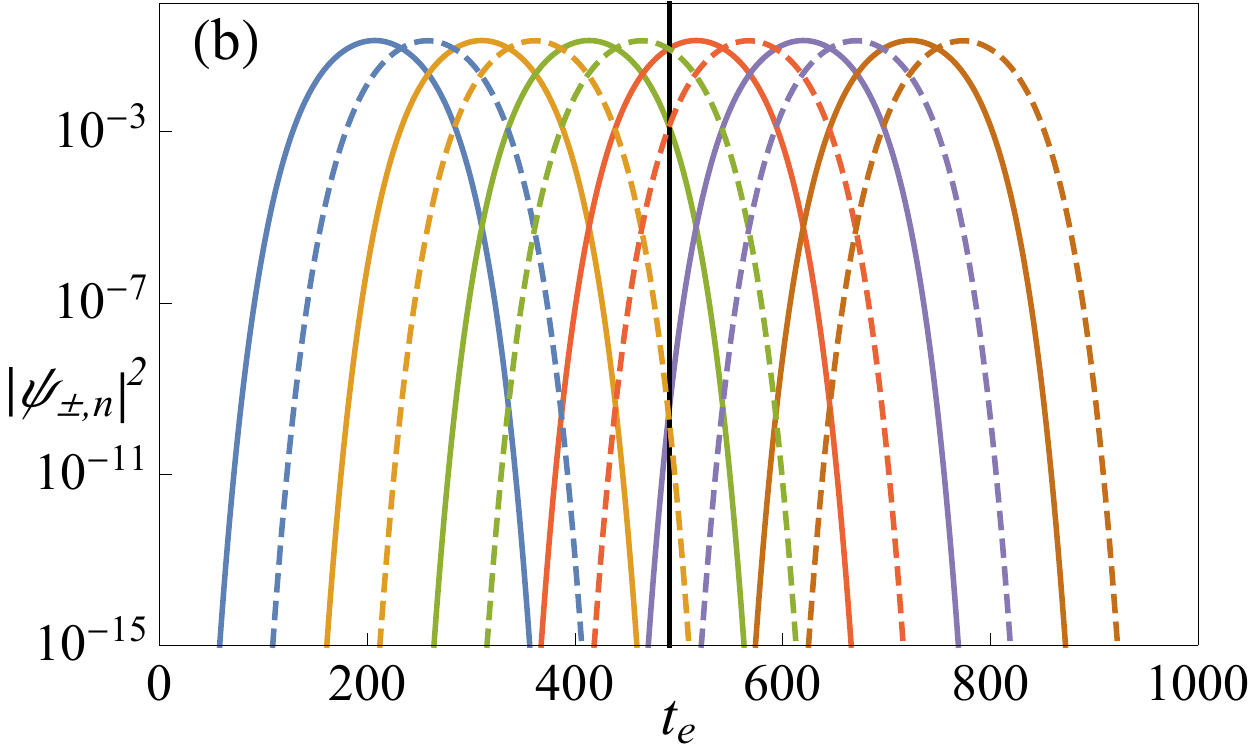}
  \includegraphics[width=0.9\columnwidth]{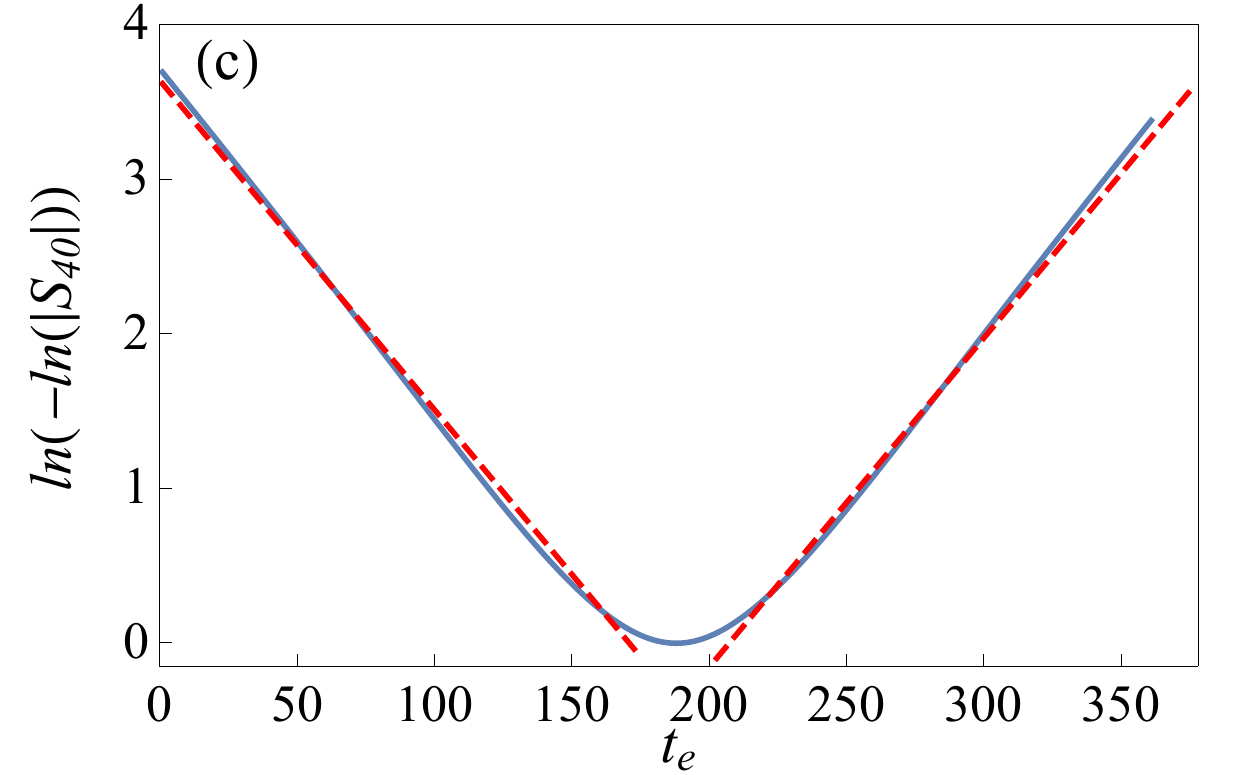}
    \caption{(Color online) 
Analysis of a moving solitary excitation with total norm $S=0.175$ and $\lambda=0.1$.
(a) Snapshot of the norm density distribution $S_n(t)$ versus $n$. 
(b) The time evolution of $|\psi_{\pm,n}|^2$  for $n=40,42,44,46,48,50$ (from left to right). 
Solid lines - $|\psi_{+,n}|^2$, dashed lines - $|\psi_{-,n}|^2$. 
The vertical black line guides the eye for the observation of a symmetry between both (see text for details).
(c) $S_{40}$ versus even time $t_e$ on double logarithmic scale (blue solid line).  Red dashed lines show the predicted superexponential decay (see Eqs. (\ref{Soliton-Shape-Sol}) and (\ref{Soliton-decay})).
}
\label{fig3}
\end{figure}

An example is shown in Fig. \ref{fig3}. It corresponds to an originally found moving solitary object with approximate speed $v\approx 1/37$
indicated by the arrow in Fig.\ref{fig2}.
The profile of the norm density distribution $S_n(t_c)$ is shown in Fig.\ref{fig3}(a). It is almost compactly localized. To zoom into the tails, we plot
$\log_{10} S_n(t_c)$ in the inset of Fig. \ref{fig3}(a) and observe tails which are decaying faster than exponential, i.e. superexponential (see more details below).
Due to the small velocity of that solitary excitation it is much more instructive to observe the evolution by following the time dependence of the wave function amplitude at a given site.
Using the SSF symmetry, we will monitor the evolution of 
$\hat \psi_n$ for a few particular even sites at even times. 
The time-dependence of $|\psi_{\pm,n}|^2$ for  the right-moving excitation from Fig.\ref{fig3}(a) is shown in Fig. \ref{fig3}(b)  for six consecutive even sites 
$n=40,42,44,46,48,50$ 
and for even times $t=2t_e$. We observe equidistantly shifted curves for each of the wave function components which indicate a motion at constant speed $v$,
and a symmetry between the evolution of
both wave function components:
\begin{equation}
\psi_{\pm,n+2}(t+\frac{2}{v})=\psi_{\pm,n}(t) \;,\;
\psi_{+,n}(t+\frac{1}{v})=-\psi_{-,n}(t)\;.
\label{symmetry}
\end{equation}
Together with faster than exponential tail decay it follows that
the rear tail profile of that right moving solitary excitation satisfies the inequalities
\begin{equation}
|\psi_{+,n}(t) | \ll |\psi_{-,n}(t)| \ll |\psi_{+,n+2}(t)| \;.
\label{Inequalities}
\end{equation}
Note that it is straightforward to generalize these inequalities to the front tail of a right moving solitary excitation, and to left moving excitations as well.
We will use these conditions below to obtain analytical results.
Finally, we plot the time dependence of the norm $S_{40}(t_e)$ in Fig.\ref{fig3}(c) on a double logarithmic scale. The superexponential decay of both the front and the
rear tails are clearly observed.

\subsection{Tail analysis}

In order to describe the tails of a moving solitary excitation, 
we expand Eq.(\ref{DiffEquation}) to first order in  $\lambda S_n \ll 1$:
\begin{equation}
\psi_{\pm,n}(t+1) =-\lambda S_{n \mp 1}(t)  \psi_{\pm,n \mp 1} (t)
\pm \psi_{\mp ,n \mp 1} (t) \;.
\label{DiffEquation-FB}
\end{equation}
Reducing the analysis to even times and sites (SSF symmetry) and accounting for the inequalities (\ref{Inequalities}) in the rear tail 
we arrive at
\begin{eqnarray} 
\psi_{+,2n}(2t) + \psi_{+,2n}(2t-2)=-\lambda G\left[ \psi_{-, 2n}(2t-2) \right]  ,
\nonumber
\\
\label{DiffEquation-soliton-FB}
\\
\nonumber
\psi_{-,2n}(2t) + \psi_{-,2n}(2t-2)=\lambda G\left[ \psi_{+, 2n+2}(2t-2) \right] ,
\end{eqnarray}
where $G\left[ \psi \right]=|\psi|^2\psi$.
To solve the above equations we use the traveling wave ansatz
\begin{eqnarray} 
\nonumber
\psi_{+,2n}(2t)=(-1)^{t+n} g_{r}(2vt-2n), 
\\
\label{Soliton-f-ansatz}
\\
\nonumber
 \psi_{-,2n}(2t)=(-1)^{t+n+1}g_{r}(2vt-2n-m),
\end{eqnarray}
with a yet to be determined argument shift $m$.
The real-valued
rear tail function $g_{r}(x)$ satisfies the difference equations
\begin{equation} \label{Soliton-moving}
 \begin{array}{cc} 
g_r(y)-g_r(y-2v) =-\lambda g_r^3(y-m), \\ \\
g_r(y-m)-g_r(y-2v-m) =-\lambda g_r^3(y-2), 
\end{array} 
\end{equation}
with $y=2vt-2n$. Since both equations have to deliver the same solution, we conclude that $m=1$, confirming the validity of Eq. (\ref{symmetry}).

If the velocity of the moving solitary excitation with core position $y=n_c$ is small, i.e. $v \ll 1$, we can replace the difference equations (\ref{Soliton-moving}) by 
a nonlinear differential equation with a discrete delay to describe the rear tail dynamics:
\begin{equation} \label{Soliton-Shape}
 2vg_r^\prime (y)=-\lambda g_r^3(y-1)\;, \; y \ll n_c \;.
\end{equation}
A straightforward generalization to the front tail dynamics yields
\begin{equation} \label{Soliton-Shape-rear}
 2vg_f^\prime (y)=\lambda g_f^3(y+1) \;, \; y \gg n_c \; .
\end{equation}
Both rear and front tail differential equations (\ref{Soliton-Shape}),(\ref{Soliton-Shape-rear}) yields the super-exponential decay solution
\begin{equation} \label{Soliton-Shape-Sol}
 g(y)=A\exp \left [-\alpha e^{|y|/\xi} +\beta |y| \right ] \;, \; |y-n_c| \gg 1 \; .
\end{equation}
Indeed, substituting  (\ref{Soliton-Shape-Sol}) into ( \ref{Soliton-Shape}),(\ref{Soliton-Shape-rear}) we obtain the super-exponential decay length $\xi$: 
\begin{equation} \label{Soliton-decay}
 \xi=1/ \ln 3 \;,\; 
\beta=1/(2\xi) \;, \; 2v \alpha/ \xi= \lambda e^{-3 \beta} A^2 \;.
\end{equation}
The super-exponential decay is plotted in Fig.\ref{fig3}(c) and agrees very well with the numerically observed front and rear tails.

\begin{figure}
\includegraphics[width=0.95\columnwidth]{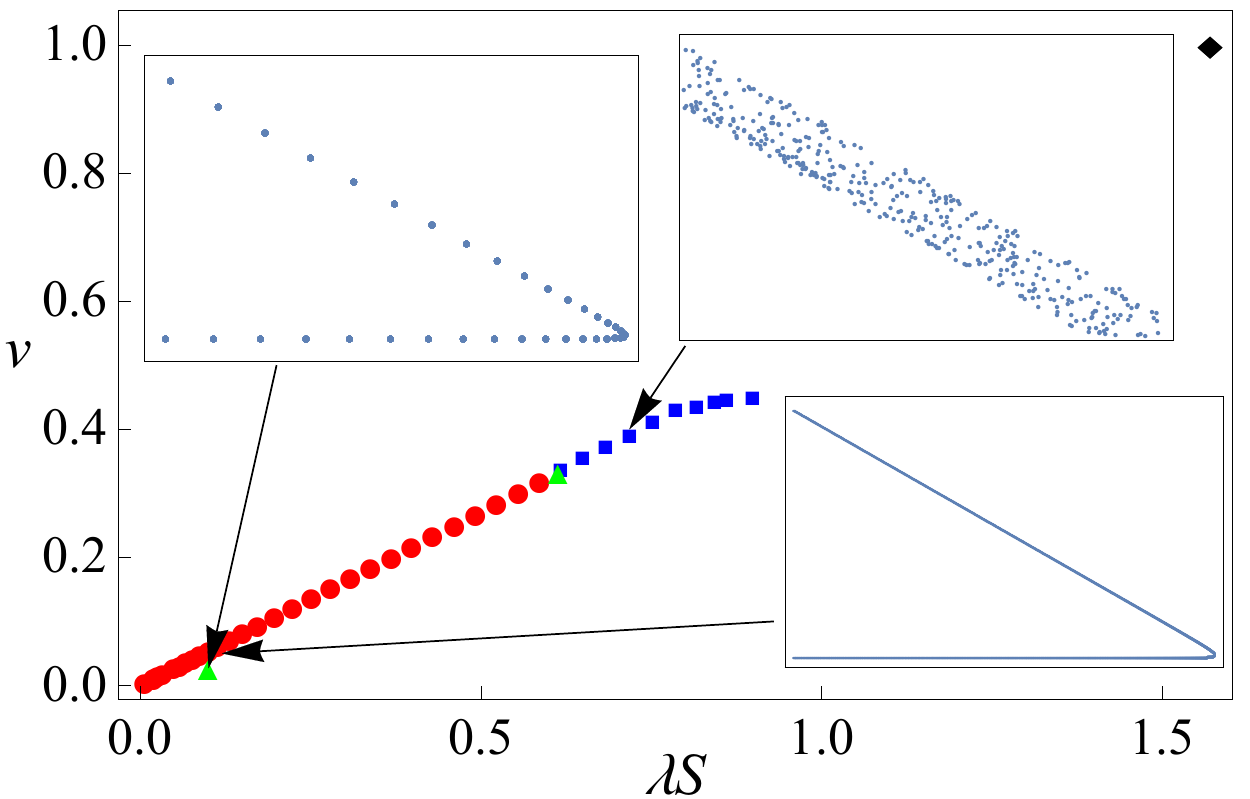}
\caption{(Color online) The dependence of the discrete breather velocity $v$ on the parameter $\lambda S$. 
Green triangles: Newton-generated periodic solutions; red circles: quasi-periodic solutions; blue squares - transient chaotic solutions. 
Black diamond - analytical bullet solution. Insets show the corresponding generalized Poincare sections. }
\label{fig4}
\end{figure}

Both tail solutions have to be glued together in the core of the moving excitation, where the above analysis does not hold.
Therefore in general $A$ could be different in front and tail for asymmetric profiles. However the numerical results  in Fig.\ref{fig3}(b,c) clearly indicate that the
 profiles are symmetric. Together with the reasonable assumption that the square amplitude in the tails is proportional to the total norm $S$ of the
whole moving solitary excitation $|A|^2 \propto S$, we arrive at the scaling relation
\begin{equation}
v \propto \lambda S \;.
\label{scaling}
\end{equation}
To test this scaling relation, we perform a single 'cut and paste' procedure to a number of moving solitary excitations as shown e.g. in Fig.\ref{fig2}, measure
their norm and velocity, and plot the result in Fig.\ref{fig4}. We observe good agreement with (\ref{scaling}) for $ \lambda S \leq 0.7$ and corresponding
velocities $v \leq 0.4$.

\subsection{Moving discrete breathers}

\begin{figure}
\includegraphics[width=0.95\columnwidth]{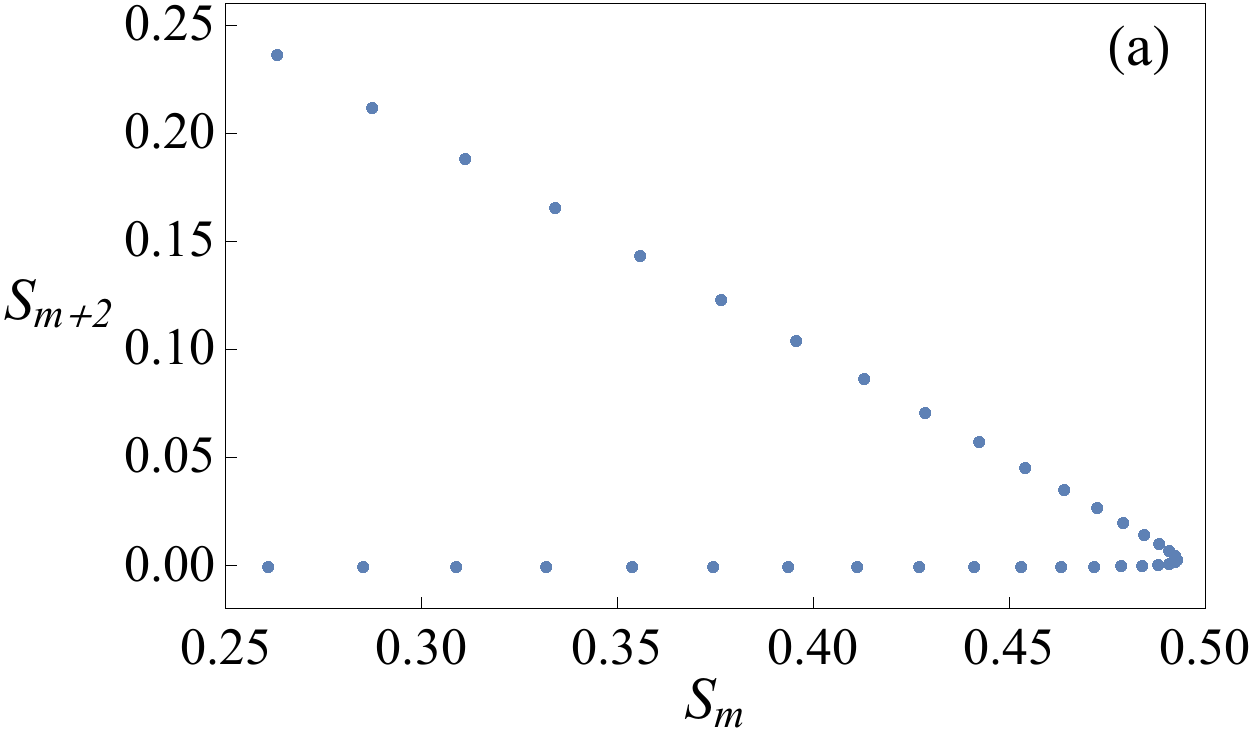} 
\includegraphics[width=0.95\columnwidth]{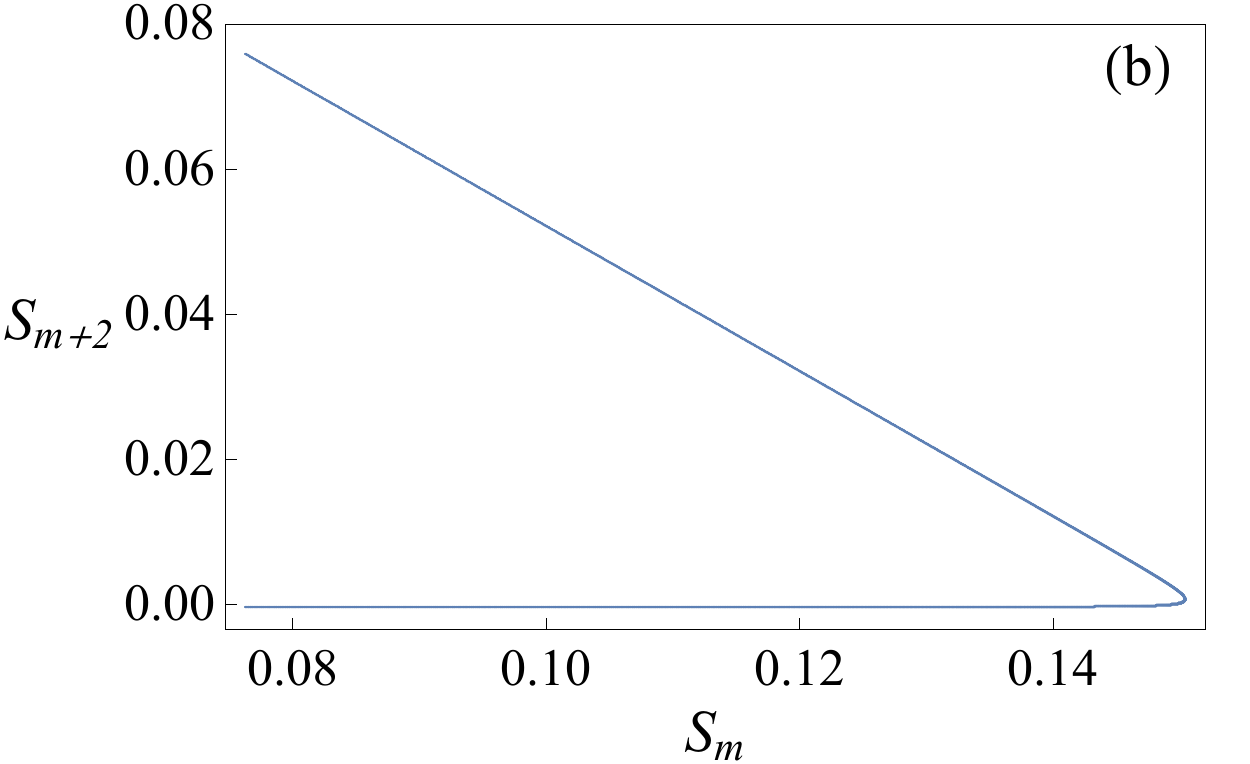} 
\includegraphics[width=0.95\columnwidth]{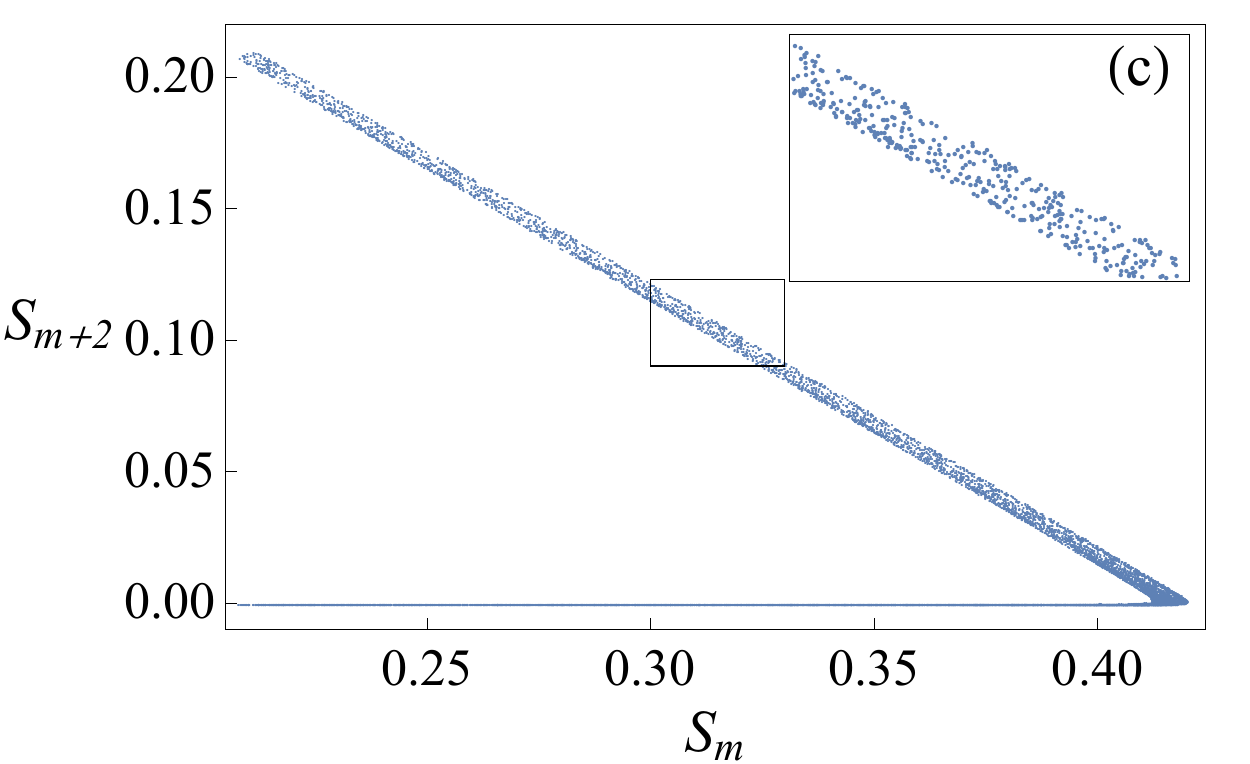}
\caption{(Color online) The Poincare sectioning of moving discrete breathers (see text for details).
(a) Periodic discrete breather with rational velocity $v=1/37$, $\lambda=0.1$ and $S=1$.
(b) Quasiperiodic discrete breather with velocity $v \approx 0.041$,  $\lambda=0.5$ and $S=0.15$.
(c) Chaotic discrete breather with $v \approx 0.39$,  $\lambda=1.7$ and $S=0.42$.
}
    \label{fig5}
\end{figure}

For the range of parameters $\lambda S < 0.9$ we were able to obtain a number of moving discrete breathers applying once the procedure of 'cut and paste'.
We characterize the internal dynamics of these objects by computing a generalized Poincare section in a co-moving frame. 
At each time $t$ we obtain the position $m$ of the largest value of $S_n(t)$ and plot $S_{m+2}$ versus $S_m$.
The results are shown in Fig.\ref{fig5}.
We find three different types of discrete breathers: periodic, quasiperiodic, and chaotic. 

Periodic moving discrete breathers are characterized by
a rational value of their velocity which leads to a finite number of points on the Poincare section as shown in Fig.\ref{fig5}(a), where $v=1/37$, $\lambda=0.1$ and $S=1$. The profile
of the moving discrete breather is fully restored after 37 iterations and one additional shift along the lattice (up to a global phase).

Quasiperiodic moving discrete breathers have an irrational velocity with a Poincare section which
forms a dense one-dimensional line segment for an infinite number of iterations. An example is shown in Fig.\ref{fig5}(b) for $v \approx 0.041$,  $\lambda=0.5$ and $S=0.15$. 

Chaotic moving discrete breathers generate a Poincare section which corresponds to a stripe segment with finite width and additional fine (potentially
fractal) structure
inside. An example is shown in Fig.\ref{fig5}(c) for $v \approx 0.39$,  $\lambda=1.7$ and $S=0.42$.
Chaotic moving discrete breathers cannot be cast into the form of a traveling wave solution (\ref{Soliton-f-ansatz}), and therefore cease to be exact solutions.
Instead, these objects are slowly losing the norm by radiating plane waves in their wake. These leftovers do not propagate further due to the flat band structure of
the small amplitude equations. Thus the chaotic moving breathers are slowing down their speed $v$. The rate of that process is probably related to the thickness
of the above stripe segments in the Poincare sections. Still, that rate can be very small, such that we observe traveling chaotic breathers over several thousands
of lattice sites without a notable change of their speed. 

Quasiperiodic moving discrete breathers instead can very well correspond to exact traveling wave solutions. However, it may well be that these objects are chaotic breathers with very narrow and thus undetected finite width of the stripe segments. It may also well be that these objects are in fact
exact traveling waves but with rational values of their velocity $v$ which lead to a period which is of the order of the simulation time.
We are not aware of fine computational means to tell these different scenarios apart.

\begin{figure}
\includegraphics[width=0.95\columnwidth]{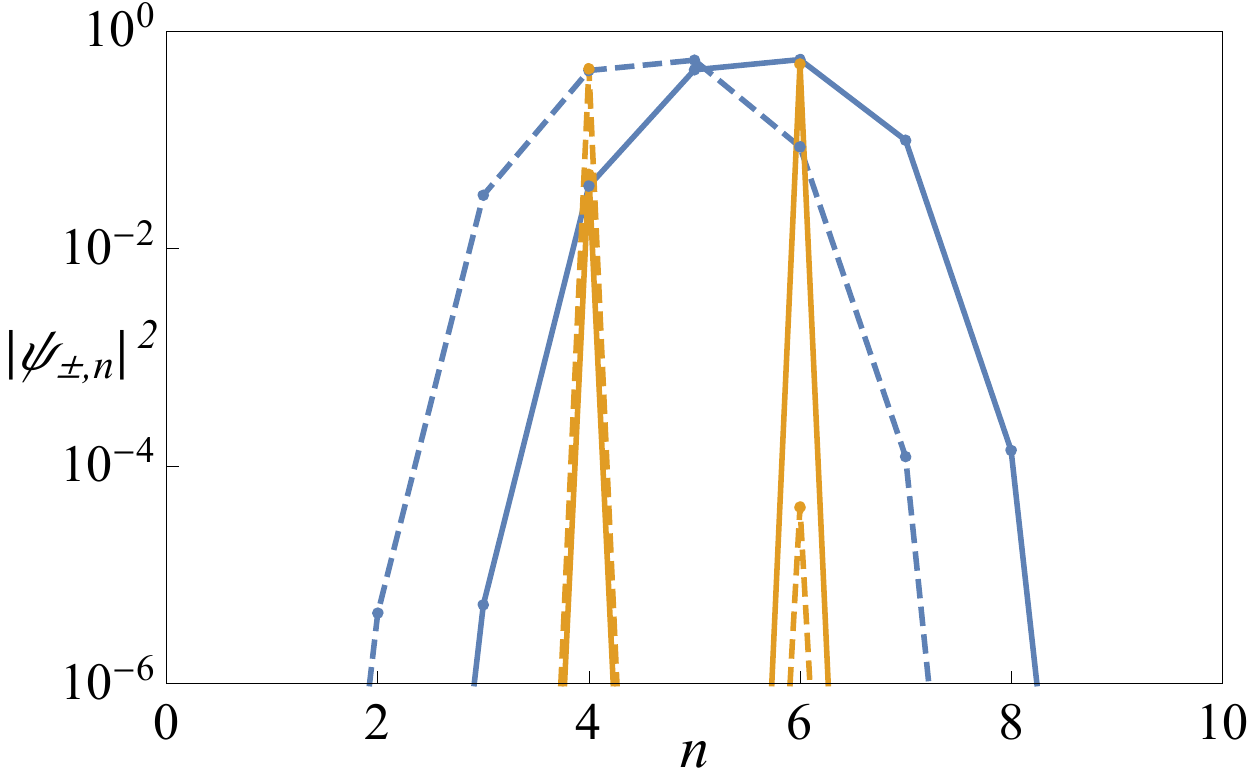} 
\caption{(Color online)  Snapshots of spatial distributions of $|\psi_\pm|^2$ for two periodic moving discrete breathers are  shown. Solid lines - $|\psi_{+n}|^2$, dashed  lines - $|\psi_{-n}|^2$. These periodic  solitary excitations correspond  to values of $p=1,q=37$ (blue lines) and $p=1,q=3$ (orange lines). } 
    \label{fig6}
\end{figure}

Periodic moving discrete breathers can be obtained with very high numerical precision using a generalized Newton scheme. 
For that we compactly rewrite Eq.(\ref{DiffEquation}) as a unitary map of the entire field $\hat{\Psi} \equiv \left \{ \hat{\psi}_n \right \}$. The action
of one iteration in (\ref{DiffEquation}) is just a nonlinear unitary map $\hat{\Psi}(t+1)=\mathcal{U} \hat{\Psi}(t)$. A translation shift along the lattice
by one lattice site will be denoted by $\mathcal{T} \left \{ \hat{\psi}_n \right \} = \left \{ \hat{\psi}_{n+1} \right \}$. Then a periodic moving breather is
encoded by two integers $p,q$ and its frequency $\Omega$ such that the breather field satisfies
\begin{equation}
\left( \mathcal{U}^q - {\rm e}^{i\Omega q} \mathcal{T}^p \right) \hat{\Psi} = 0 \;, \; v = \frac{p}{q} \;.
\label{Newton}  
\end{equation}
Solutions to (\ref{Newton}) can be obtained using standard Newton schemes which search for zeros of vector functions \cite{Press:2007aa}.
We found two such solutions. The first one was described above, and has velocity $v=1/37$, $q=37,p=1$ and $\Omega=\pi/37$.  The second one has
velocity $v=1/3$, $q=3,p=1$ and $\Omega = 2\pi/3$. Both solutions are shown with green circles in Fig.\ref{fig4}. The profiles of both solutions are shown
in Fig.\ref{fig6}.

\subsection{Bullets}
\begin{figure}
\includegraphics[width=0.95\columnwidth]{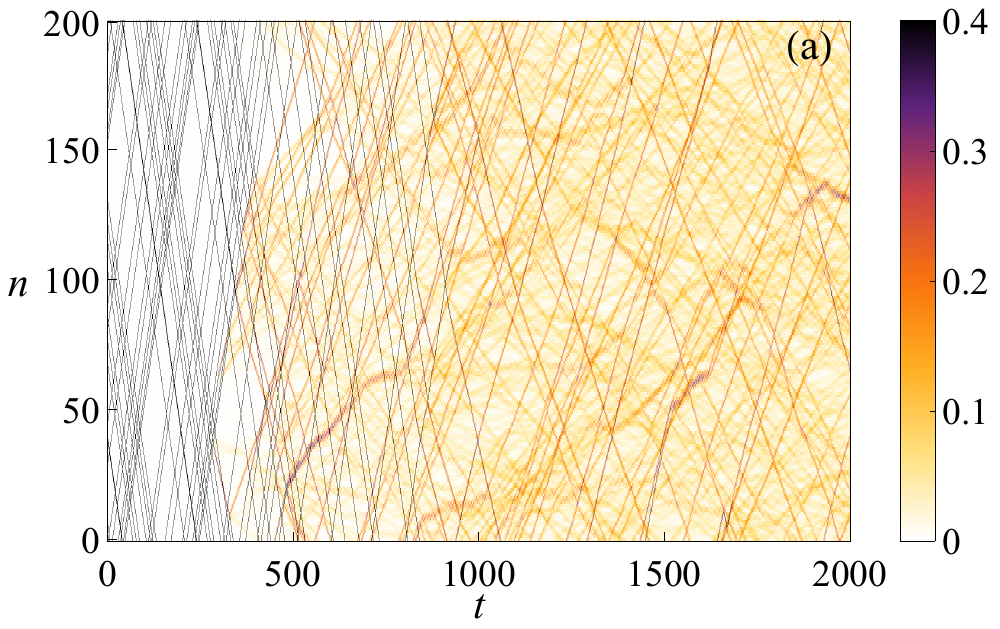} 
\includegraphics[width=0.95\columnwidth]{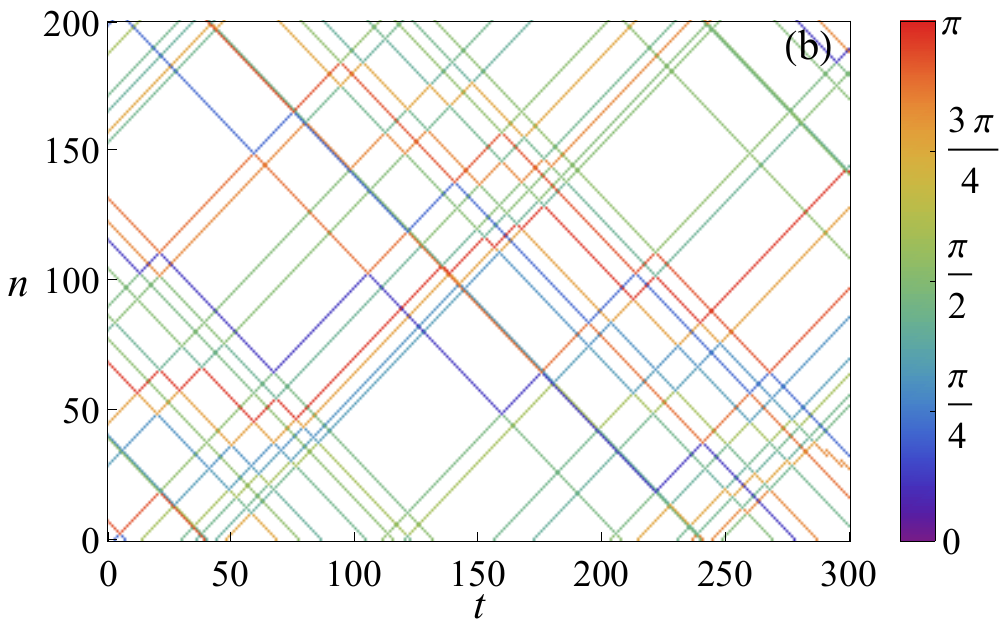} 
\includegraphics[width=0.95\columnwidth]{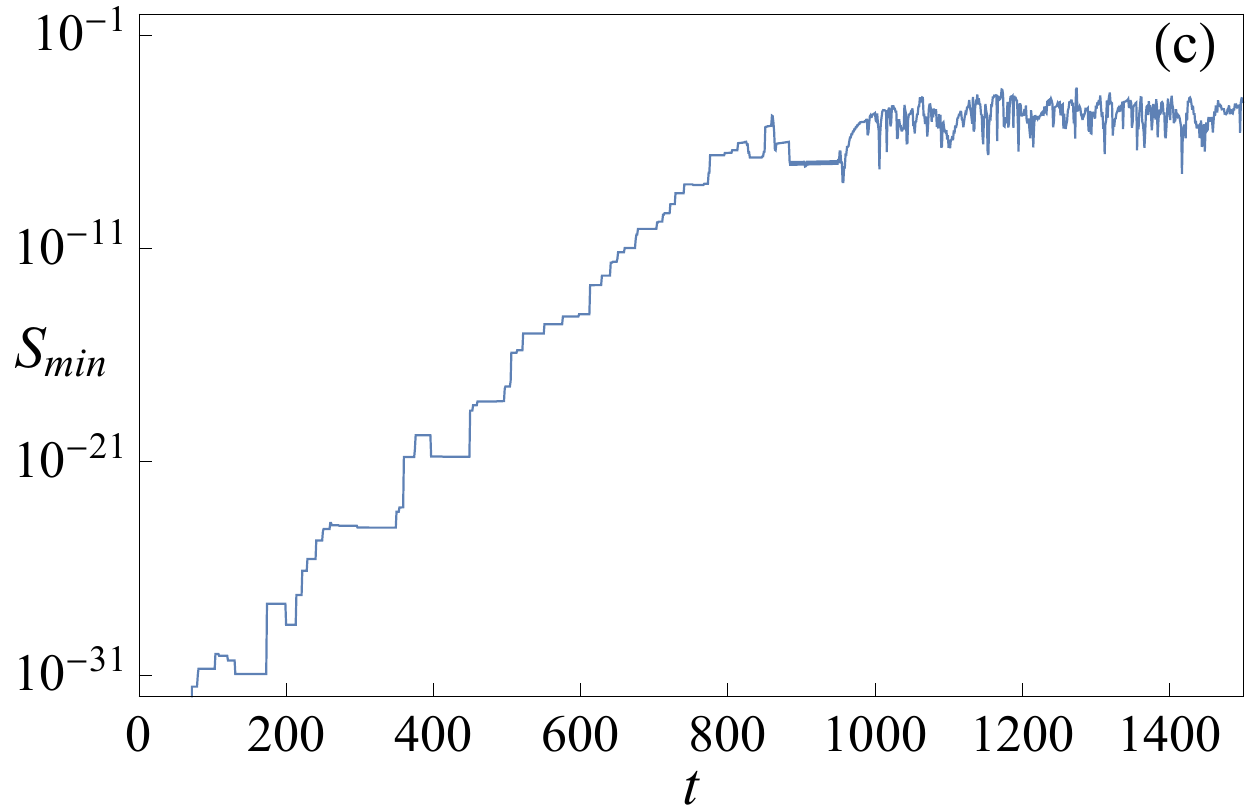}
\caption{(Color online)  Evolution of a bullet gas with randomly chosen initial coordinates and phases. (a) Spatial and temporal dependence of the norm density, $S_n(t)$; (b) Bullet phases (modulo of $2\pi$) , $\phi_{\pm}$, as function of space and time; (c) Time evolution of the smallest $S_{min}$.   }
\label{fig7}
\end{figure}

For the particular value of the nonlinearity $\lambda S=\pm \pi/2$ the equations (\ref{DiffEquation}) admit exact moving and compact excitations, which we coin {\sl bullets}.
These bullets have a nonzero field amplitude on only one lattice site and in addition on only one of the two wave function components. Its velocity $v=\pm 1$
(see Fig.\ref{fig4}).
A right-moving bullet is given by 
\begin{equation}
\hat{\psi}_n(t) = (-1)^{s t} \sqrt{S} 
\left \{
{\rm e}^{i\phi_+},0
\right \}
\delta_{n,n_0+t}\;,
\label{rightbullet}
\end{equation}
with $s=\pm1$ being the sign of $\lambda$. A left-moving bullet respectively is given by
\begin{equation}
\hat{\psi}_n(t) = (-1)^{s t} \sqrt{S} 
\left \{0,
{\rm e}^{i\phi_-}
\right \}
\delta_{n,n_0-t} \; .
\label{leftbullet}
\end{equation}
Thus, the dynamics of a single bullet is characterized by two conserved quantities - the total norm $S=\pi/(2|\lambda|)$ and the phase $\phi_{\pm}$. Interestingly, the dynamics of \textit{a gas} of left and right moving bullets shows the conservation of  the numbers of left and right movers. Two bullets approaching each other will experience no interaction if their relative distance is odd, but will undergo an act of elastic scattering if that distance is even. After such
an elastic reflection a left mover is turned into a right mover (carrying its phase $\phi$ with it)  and vice versa. The evolution of such a gas of left and right moving bullets is shown in Fig.\ref{fig7}(a,b). 
The dynamics of the gas will lead to instabilities in the presence of a small noisy background.
This background is generated during the computation due to roundoff errors. It is observed in Fig.\ref{fig7}(a) at times $t \approx 500$. The time dependence
of the smallest background norm $S_{min}$ during that evolution is shown in Fig.\ref{fig7}(c). The background intensity is growing exponentially fast up to times
$t\approx 1400$. Yet for larger times the system still shows up with an interacting gas of moving solitary excitations with various velocities, see Fig.\ref{fig7}(a).

\subsection{Stationary discrete breathers}

\begin{figure}
\includegraphics[width=0.95\columnwidth]{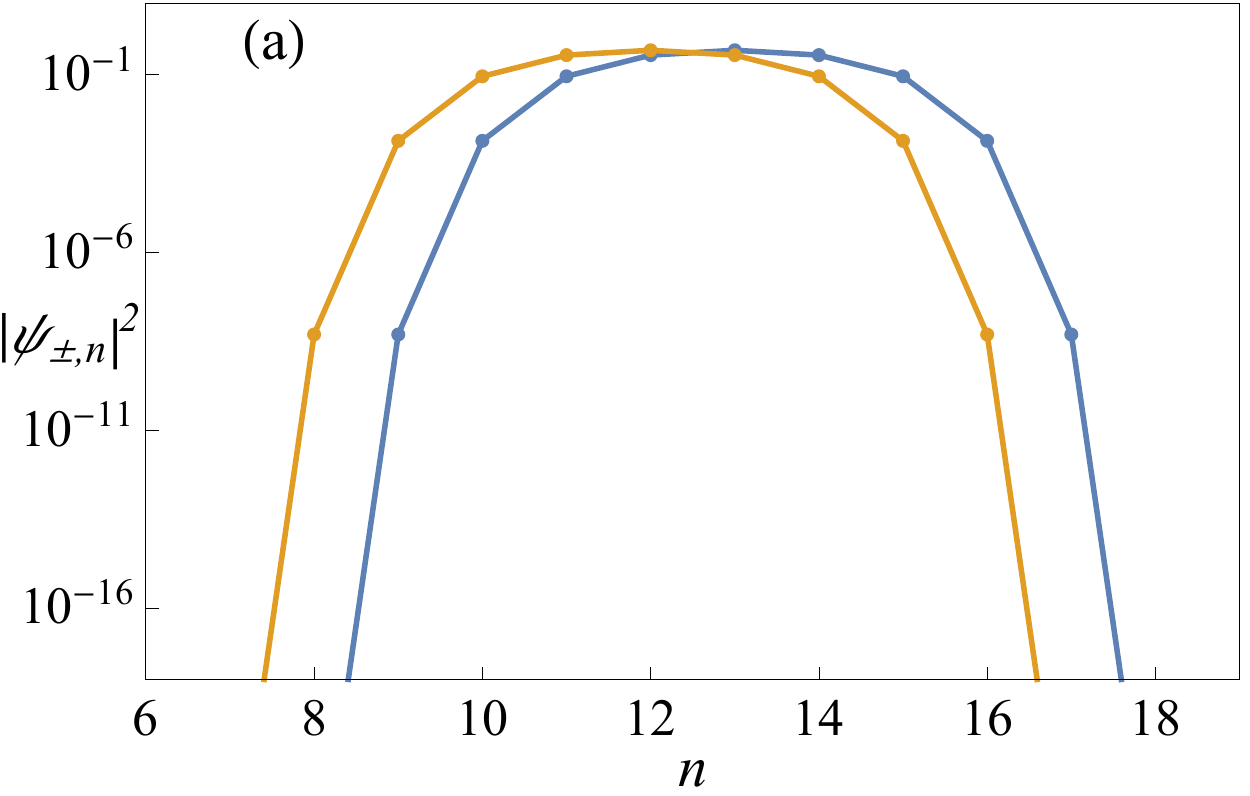} 
\includegraphics[width=0.95\columnwidth]{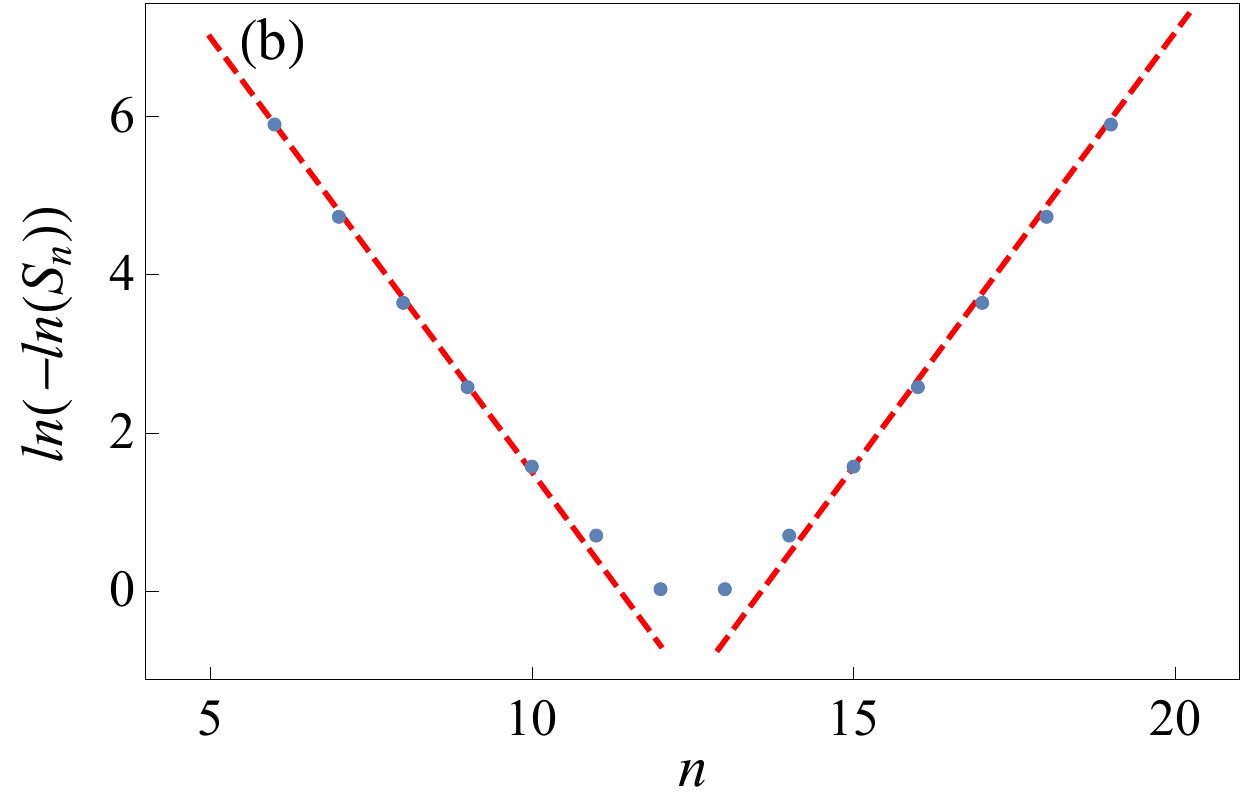}
\includegraphics[width=0.95\columnwidth]{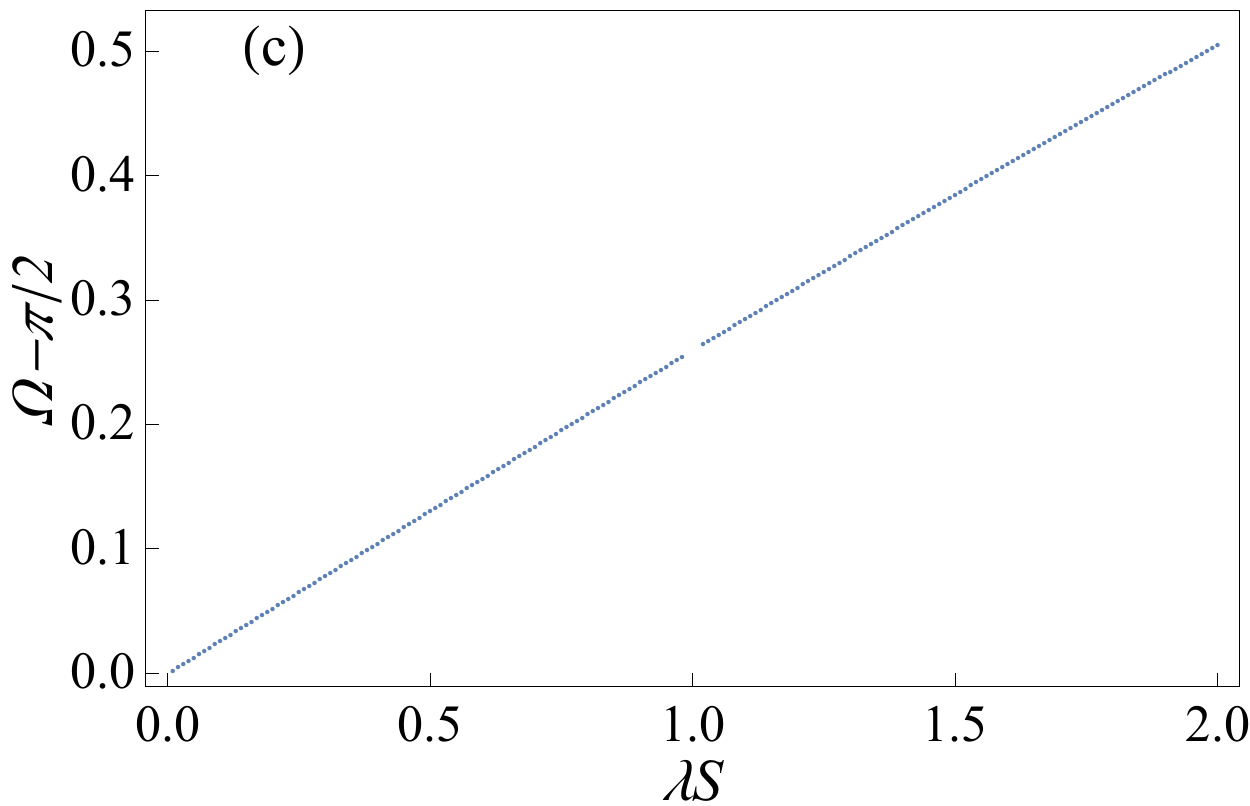}
\caption{(Color online) 
(a) A snapshot of a single frequency stationary breather solution. Solid line - $|\psi_{+n}|^2$, dashed line - $|\psi_{-n}|^2$; (b) Spatial dependence of $S_n$ showing  super-exponential decay in its tails (the red dashed line corresponds to the solution of (\ref{Breather}); (c) Scaling  of the breather frequency with the parameter $\lambda S$. }
    \label{fig8}
\end{figure}

In addition to the above surfeit of moving discrete breathers, we also report on the existence of stationary discrete breathers with zero velocity. These objects are zeros
of the nonlinear map
\begin{equation}
\left( \mathcal{U}^q - {\rm e}^{i\Omega q} \mathcal{I} \right) \hat{\Psi} = 0  \;,
\label{SNewton}  
\end{equation}
where $\mathcal{I}$ is the unity operator. 
Below we will consider the case $q=1$ only.
Solutions can be again searched for by using a generalized Newton scheme. 
The spatial profile of one of these solutions is shown in Fig.\ref{fig8}(a). A double logarithmic plot of the profile in
Fig.\ref{fig8}(b) reveals its superexponential tails. Interestingly, the deviation of the stationary breather frequency $\Omega$ from $\pi/2$ yields
a linear dependence on the control parameter $\lambda S$ (see Fig.\ref{fig8}(c)).

To obtain an analytical solution for the tails of a stationary breather, we use the observation from Fig.\ref{fig8}(a) that one of the two components is dominating in the tail, e.g.
$|\psi_{+,n}| \gg |\psi_{-,n}|$ for the right tail of the breather. Note that the final result will be invariant on the choice of the tail.
With the ansatz
\begin{equation} \label{DB-ansatz}
 \begin{array}{cc} 
\psi_{+,n}(t)=\exp[i\Omega t+\phi_+] g(n), \\ \\ \\ 
\psi_{-,n}(t)=\exp[i\Omega t+\phi_-] g(n+n_0), \\
\end{array} 
\end{equation} 
and Eq. (\ref{DiffEquation-FB}) we obtain $n_0=1$ and 
\begin{equation} \label{Breather}
2g(n) \cos \Omega =-\lambda g^3(n-1).
\end{equation} 
The solution of the above nonlinear difference equation yields the super-exponential decay
$ g(n)=A\exp[-\alpha e^{|n|/\xi}],~~|n| \gg 1$ with $\xi=1/\ln 3$. 
The frequency of the breather is determined by both the nonlinearity and its norm:  $(\Omega-\pi/2) \propto \lambda S$, and this scaling is in a good accord with numerical analysis (see the Fig. \ref{fig8}(c)). 

\section{Conclusions}
\label{IV}
Discrete time quantum walks are unitary maps defined on the Hilbert space of coupled two-level systems, which turn into a very efficient unitary toolbox for addressing a variety of problems
which lack easy solutions in Hamiltonian settings.
In particular, we studied the dynamics of excitations in a nonlinear discrete time quantum walk, 
whose fine-tuned linear counterpart has a flat band structure. 
The linear counterpart is, therefore, lacking transport, with exact solutions being compactly localized. A solitary 
entity of the nonlinear walk moving at velocity $v$ is then shown 
to not
suffer from resonances with small amplitude plane waves with identical phase velocity, due to the absence of the latter. 
That solitary excitation also shows
to be localized stronger than exponential, due to the absence of a linear dispersion.
We found a set of stationary and moving breathers with almost compact superexponential spatial tails. 
At the 
limit of the largest velocity $v=1$ the moving breather turns into a completely compact bullet.  
Remarkably these bullets can form an interacting gas
with the bullet phases participating in the scattering process. 
It remains an interesting question as to whether such highly localized 
- and even compact - moving nonlinear solitary excitations may be used in applications involving discrete time quantum walks.

\textbf{Acknowledgements}
This work was supported by the Institute for Basic Science in Korea (IBS-R024-D1). 
M. V. F. acknowledges the partial support of the Russian Science Foundation (grant No. 16-12-00095).
Y.Z. acknowledges the support of the National Academy of Sciences of
Ukraine through program No. 0117U000236.

\bibliographystyle{unsrt}
\bibliography{main}
\end{document}